\input harvmac
\def \h {\hat}

\def \s {\sigma}
\def \p {\phi}
\def \h {\hat }

\def \ha {\half}
\def \ov {\over}

\def \four{{\textstyle {1\ov 4}}}
\def \a {\alpha}
\def \lr { \lref}
\def\ep{\epsilon}

\def \r {\rho}
 \def\m{\mu}\def\n {\nu}\def\l
{\lambda}
\def \del {\partial}

\def \ha{{\textstyle{1\over 2}}}

\def \a {\alpha}
\def \b {\beta}
\def \zeta {\zeta}
\def \s {\sigma}
\def \p {\phi}
\def \m {\mu}
\def \n {\nu}

\def \t {\theta}

\def \sm {$\s$-model\ }
\def \P {\Phi}

\def \ov {\over }
\def \four{{\textstyle{1\over 4}}}

\def \third {{\textstyle{1\over 3}}}
\def\g {\gamma}
\def \P {\Phi}

\def \k {\kappa}
\def\np {  Nucl. Phys. }
\def \pl { Phys. Lett. }
\def \mpl { Mod. Phys. Lett. }

\def \pr  { Phys. Rev. }
\def \cqg { Class. Quantum Grav.}

\baselineskip8pt
\Title{\vbox
{\baselineskip 6pt{\hbox{  }}{\hbox
{Imperial/TP/95-96/19  }}{\hbox{hep-th/9601109}} {\hbox{
   }}} }
{\vbox{\centerline {On dilaton dependence of type II superstring action }
 \centerline {  }
 }}





\medskip
\centerline{   A.A. Tseytlin\footnote{$^{\star}$}{\baselineskip8pt
e-mail address: tseytlin@ic.ac.uk}\footnote{$^{\dagger}$}{\baselineskip8pt
On leave  from Lebedev  Physics
Institute, Moscow.} }

\smallskip\smallskip
\centerline {\it  Theoretical Physics Group, Blackett Laboratory,}
\smallskip

\centerline {\it  Imperial College,  London SW7 2BZ, U.K. }
\bigskip\bigskip
\centerline {\bf Abstract}
\medskip
\baselineskip10pt
\noindent
\medskip
The supersymmetric action of  type IIA $D=10$ superstring in $N=2a,\  D=10$
supergravity background can be derived by double dimensional reduction 
of the action of supermembrane coupled to $D=11$ supergravity. We 
demonstrate that the background Ramond-Ramond fields appear in the
resulting superstring action with an extra factor of exponential of 
the dilaton.

\Date {January 1996}

\noblackbox
\baselineskip 16pt plus 2pt minus 2pt

\def \k {\kappa}

\lr \chap {A.H. Chamseddine, \np B185 (1981) 403;
E. Bergshoeff, M. de Roo, B. de Wit and P. van Nieuwenhuizen, \np B195 (1982)
97;
G.F. Chaplin and N.S. Manton, \pl B120 (1983) 105.}

\lr \witten {E. Witten, \np B443 (1995) 85, hep-th/9503124. }

\lr \ft { E.S. Fradkin  and A.A. Tseytlin, \pl B160 (1985) 69.  }

\lr \ftt { E.S. Fradkin  and A.A. Tseytlin, \pl B158 (1985) 316;  
 A.A. Tseytlin,  \pl B208 (1988) 221. }
\lr \tsee {Tseytlin }

\lr \gs{ M.B. Green  and J.H. Schwarz, \pl B149 (1984) 117;
\pl B151 (1985) 21; \np B255 (1985) 93. }

\lr \callan {C.G. Callan, C. Lovelace, C.R. Nappi and S.A. Yost, \pl B206
(1988) 41;
\np B308 (1988) 221.}
\lr \green {M.B. Green, J.H.  Schwarz and E.  Witten, {\it Superstring Theory}
(Cambridge U.P., 1988).}

\lr \green {M.B. Green, J.H.  Schwarz and E.  Witten, {\it Superstring Theory}
(Cambridge U.P., 1988).}

\lr \nils{ B.E.W. Nilsson and A.K. Tollsten, \pl 169 (1986) 369; R. Kallosh, 
Phys. Scr. T15 (1987) 118.  }
\lr \tsss{ } 
\lr \alw{L. \'Alvarez-Gaum\'e and E. Witten, \np B234 (1983) 269.     }
\lr \crem { E. Cremmer  and S. Ferrara, \pl B91 (1980) 61. }
\lr \bri { L. Brink  and P. Howe, \pl B91 (1980) 384. }
\lr \duf { M.J. Duff, P.S. Howe, T. Inami and K.S. Stelle, 
\pl B191 (1987) 70. }

\lr \pol {J. Polchinski, ``Dirichlet-branes and Ramond-Ramond charges", NSF-ITP-95-122, hep-th/9510017.}
\lr \doug { Douglas}
\lr\poly{D. Polyakov,  ``RR - dilaton interaction in type II superstring",
RU-95-85,  hep-th/9512028.}
\lr \hul{E. Bergshoeff, C.  Hull  and T. Ortin, \np  B451 (1995) 547, hep-th/9504081.}
\lr\gups{Gupser}
\lr \seza {M. Huq and M.A. Namazie, \cqg 2 (1985) 293, 597 (E); 
F. Giani and M. Pernici, \pr D30 (1984) 325;
I.C. Campbell and P.C. West, \np B243 (1984) 112;
S.J. Gates, Jr., J. Carr and R. Oerter, \pl  B189 (1987) 68.}
\lr \sezb{
J.H. Schwarz, \np B226 (1983) 269;
P.S. Howe and P.C. West, \np B238 (1984) 181.}

\lr \gs{ M.B. Green and J.H. Schwarz, \pl B136 (1984) 376; \np B243 (1984) 285.}

\lr \wiit{E. Witten, \np B266 (1986) 245; 
J.J. Atick, A. Dhar and B. Ratra, \pl B169 (1986) 54;
R. Kallosh, Phys. Scr. T15 (1987) 118;
M.T. Grisaru and D. Zanon, }

\lr \mizi{M.T. Grisaru, P.S. Howe, L. Mezincescu, B.E.W. Nilsson and P.K. Townsend, \pl B162 (1985) 116.  }
\lr \achu {A. Ach\'ucarro, P. Kapusta and K.S. Stelle, \pl B232 (1989) 302.}
\lr \berg{ E. Bergshoeff, E. Sezgin and P.K. Townsend, \pl B189 
(1987) 75; Ann. of Phys.  185 (1988) 330. }

\lr \pkt {P.K. Townsend, \pl B350 (1995) 184.}
\lr \cree{E. Cremmer, B. Julia and J. Scherk, \pl B76 (1978) 409.}
\lr \ber {N. Berkovits and W. Siegel, ``Superspace effective actions for 4D compactifications of heterotic and type II superstrings", 
IFUSP-P-1180, ITP-SB-95-41,
hep-th/9510106.}

\lr \wit{S.J. Gates,  Jr.,  and B. Zwiebach, \np B238 (1984) 99;
 E. Witten, \pl B155 (1985) 151;
R.E. Kallosh, \pl B159 (1985) 111.}
\lr \schw {J.H. Schwarz, Phys. Rep. 89 (1982) 223.}
 \lr \gatt{S. Bellucci, S.J. Gates, Jr., B. Radak, P. Majumdar
      and Sh. Vashakidze,  \mpl  A4 (1989) 1985;
S.J. Gates, Jr., P. Majumdar, B. Radak and  Sh. Vashakidze,
     \pl 
     B226 (1989) 237;  B. Radak and Sh. Vashakidze, \pl B255 (1991) 528.}

1.  Recent discussions of  string  dualities 
involve  bosonic fields which originate from the Ramond-Ramond (RR)
sector of the  superstring. 
An  important observation  \witten\  is 
  that  kinetic terms of the  (manifestly gauge-covariant) RR fields 
in the $D=10$ supergravity actions expressed  in terms of the 
string-frame metric are 
 {\it not}  multiplied by  the standard  tree-level $e^{-2\p}$ dilaton 
factor  (see also \hul).  
  It    would  be   important to  have a clear   world-sheet 
understanding of  this  fact. This would allow one,  in particular,  
 to go beyond the supergravity level 
and determine the dilaton dependence 
of  terms involving higher powers of RR fields 
in  $D=10$ superstring effective actions.

A natural conjecture is that the RR fields enter the 
background-dependent 
superstring  action with an extra  $e^{\p}$ factor in front 
of their coupling terms. Then  
the dilaton dependence cancels out   in the 
RR kinetic terms in  the tree-level partition function (which is, up to a renormalisation,  the superstring 
effective action \refs{\ftt,\ft}), 
$S= \int d^{10} x \sqrt G [e^{-2\p} (R + ...) +  F^2 + ...]$,  
and, correspondingly,  survives in the 
conformal anomaly  $\b$-function, $R  +...+  e^{2\p} F^2 + ...=0$.

The  reason for this special coupling of the RR fields to the dilaton lies in  the structure  of  local 
$N=2, D=10$ supersymmetry. 
Since  the $D=10$  supersymmetry is not manifest
in the Ramond-Neveu-Schwarz   formulation, the RR-field--dilaton coupling  is  hard to determine 
 there
 at the full non-linear level. One can still try to  explain the presence
of extra $e^{\p}$-factor
 using  linear dilaton  background \pol\ 
or suggest a heuristic argument 
relating  it 
to non-locality of  the RR vertex (spin)  operators
(which  effectively cut a hole out of the world sheet and thus change its Euler number by  $-1$) \refs{\poly}. 
It is possible, of course,  to 
compute the 3-point RR-RR-dilaton amplitudes 
to check \refs{\poly}   
that they are consistent with the structure of the  $D=10$  supergravity actions.

The   proper framework  for  addressing this question 
 should be the  
Green-Schwarz (GS) formulation  \gs. 
Starting with the free GS action one  can   couple 
it to  a  
 supergravity background in a way preserving $D=10$ supersymmetry.
Since  the $N=2$, $D=10$ supersymmetry transformation laws 
are `inhomogeneous' in  the dilaton \refs{\seza,\sezb}
(i.e. contain powers of $e^{\p}$ even for the
 `string-frame' choice of the metric)\foot{It is a peculiarity
of $N=1,D=10$ supergravity that there exists a choice 
of the metric (heterotic string frame metric) for which 
there are no $e^{\p}$ factors in the supersymmetry transformation laws, 
so that the full invariant Lagrangian 
 has $e^{-2\p}$ in front of it \wit, 
in agreement with the absence of the RR fields
 in the heterotic string spectrum. The $e^\p$ factors re-appear if one uses the 
type I string frame metric (cf. \witten).}
one should not be surprised to find the $e^{\p}$ factors 
being present in the resulting  world-sheet   superstring \sm action.

The coupling of GS  superstring to RR fields was discussed
in the light-cone gauge  
in \ft\ (starting with the  known  light-cone gauge  GS
vertex operators \schw) but since the condition of  supersymmetry 
of the action was not imposed,  
the dilaton dependence  was not determined 
(though the special way of how  RR fields appear in the space-time effective action 
was noticed and  was conjectured to be related to the fact that they couple to the fermionic part of the GS  \sm action). 
Superspace expressions for  
covariant type II  GS superstring actions in  (on-shell) 
$N=2,  D=10$ supergravity backgrounds 
were found  
  in \refs{\mizi, \duf} and studied in detail  in \gatt,  
but   component form of the actions 
(in particular, the dilaton dependence of the RR  coupling terms) 
was not explicitly worked out.
A systematic  approach to construction of manifestly $N=2, D=4$
supersymmetric world sheet \sm  (and  effective action) for 
  compactified $D=4$
Type II  superstring    was  recently 
 presented in \ber,  where a 
peculiar dilaton coupling to  the  $D=4$ 
RR  vector fields was also  discussed.

Our aim here is to  demonstrate  that there is, indeed, 
the  $e^\p$ dilaton factor in front of the RR field
coupling terms in the background-dependent 
$D=10$  type IIA  GS superstring action.
We shall follow \duf\ and derive  (the relevant terms in) the 
superstring action
by   double dimensional reduction  
of the action of  $D=11$ supermembrane  in  
$D=11$ supergravity background \berg.
Since the $N=2a, D=10$ supergravity \seza\ is a dimensional reduction 
of the $D=11$ theory \cree\ 
  this  may be viewed  just as a useful  trick to organise the computation.
 This approach may, however, provide  also    an insight
into the $D=11$ origin of these special dilaton -- RR-field  
couplings. 
As is well known, the 
 dilaton appears from the $(11,11)$ component of the  $D=11$ metric 
while the RR bosons $A_1=(A_\m)$ and $A_3= (A_{\m\n\l})$ originate from the  
$(11,\m)$ component of the metric and  the 3-form field of the 
$D=11$ supergravity.
 While the  $D=10$ metric $G_{\m\n}$
 and  the 2-form field $B_{\m\n} = A_{\m\n 11}$  have  the 
standard bosonic \sm type couplings,  we shall see that 
 the RR coupling terms involve fermionic coordinates, 
depend 
only on the gauge-invariant
field strengths
 $F_2=dA_1, \ F_4=dA_3 - 3 A_1 \wedge dB$ 
and,  as expected, contain  the extra $e^\p$ factor  
 compared  to the 
`NS-NS' coupling terms in  the  GS  \sm action.

2. As in  \duf\ our  starting point 
 is  the action of the $D=11$ supermembrane 
coupled to $D=11$ supergravity background \berg\
\eqn\mem{ I= c \int d^3 \xi [\sqrt g ( g^{ i j} \h E^m_i \h E^n_j \eta_{mn} -1)
- \third  \ep^{ijk} \h E^A_i\h E^B_j \h E^C_k A_{CBA} ] \ . }
Here  
$\h E^A_i = \del_i Z^M \h E_M^A (Z)$, \ $i=1,2,3$, \  
$Z^M=(x^\m, \t^\a)$, \ $\m= 1,..., 11$, $\a=1,..., 32$, 
\ $ A_{CBA}= A_{CBA} (Z)$, \
and $A=(m,a)$ is the corresponding tangent space index
($m=1,...,11; \ a=1,...,32$).
As  was shown in \refs{\berg,\duf}
this   action is invariant under $\kappa$-supersymmetry
provided  the background  satisfies  the superspace 
equations of on-shell $D=11$ supergravity \refs{ \crem,\bri}.
The  background superfields  have the following 
expansions \crem\ in terms of the component fields
$\{\h e^m_\m(x), \psi^a_\m(x), A_{\m\n\l}(x)\}$ of $D=11$ supergravity 
\eqn\sup{\h E^m_\m = \h e^m_\m  - i \bar \t \g^m \psi_\m + O(\t^2), \ \  }
$$
\h E^a_\m = \psi^a_\m   + 
\four  \h \omega_{\m mn}  (\g^{mn}\t)^a
+  \h F_{mnpq} \h e^s_\m (T_s^{\ mnpq} \t)^a  + O(\psi^2\t,\t^2), $$
$$\h E^m_\a = \ha i (\bar \t \g^m)_\a +  O(\t^2), \ \ \ \ 
\h E^a_\a = \delta^a_\a +  O(\t^2), $$
\eqn\aaa{ A_{mnp} (x,\t)=  A_{mnp}
 - {\textstyle {3\ov 2} } \bar \psi_{[p } \g_{mn]} \t
+  O(\t^2), \ \ \ 
A_{mna} (x,\t)= -\four (\bar \t \g_{mn})_a +  O(\t^2), } 
$$ 
A_{mab} (x,\t)= A_{abc} (x,\t)=O(\t^2),   $$ 
where
\eqn\deff{ 
\h F_{\m\n\r\l}\equiv  4 \del_{[\m} A_{\n\r\l]} = \h e^m_\m \h e^n_\n \h e^p_\r \h e^q_\l \h F_{mnpq}, \ \  \ \ T^{smnpq} \equiv {{\textstyle{1\over 144}}} (\g^{mnpqs} - 8 \g^{[mnp}\eta^{q]s}),  }
and $\h \omega_{\m mn}(\h e)$ is the standard vielbein  connection.

To relate \mem\ to  the type IIA  GS superstring  action 
in  $N=2a, D=10$ supergravity background 
we perform the double dimensional reduction  \refs{\duf,\achu}
by splitting the  world-volume and space-time 
coordinates in  `$2+1$' and `$10+1$' way,  
$\xi^i=(\xi^{i'},\xi^3), \ {i'}=1,2$, \ $x^\m=(x^{{\m'}},x^{11}\equiv y)$, 
\ ${\m'}=1,...,10$, assuming
that $\del_3 g_{ij}=0$, $\del_3 x^{\m'} =0,$ \  $\del_3 \t^\a=0$, 
$\del_{11} \h E^A_M=0, \ \del_{11} A_{MNK}=0$
and making a partial gauge choice  by relating  the `extra' world-volume and space-time coordinates, $y=\xi^3$.
It is not necessary, of course, to split the fermionic indices $(\a,a)$.
In what follows the  primed indices $\m',\n',..., $ and $m',n',...,$ will run from 1 to 10
and   
$y$  will be used to  indicate  the 11-th coordinate index.

The  bosonic  fields of the $D=11$ supergravity 
are split as follows:
\eqn\comp{ \h e^m_\m = e^{- {1\ov 3}\p} \pmatrix{e^{m'}_{\m'}  & e^\p A_{{\m'}}  \cr
0 & e^\p \cr}  
  ,  \ \ \ \ A_{\m\n\l} = \{A_{{{\m'}}\n'\l'}, \ A_{{{\m'}}\n' y}\equiv B_{{{\m'}}\n'}\} \ , }
$$ 
 \h G_{\m\n} = \h e^m_\m \h e^n_\n \eta_{mn} = e^{-{2\ov 3}  \p}
\pmatrix{G_{{{\m'}}\n'}  + e^{2\p} A_{{\m'}}A_{\n' } & e^{2\p} A_{{\m'}}  \cr
 e^{2\p} A_{\n'}& e^{2\p} \cr} .  
 $$
The dilaton factors are chosen so that they drop out of the 
bosonic part of the resulting superstring action  (after a rescaling or elimination 
of the  world-sheet metric), 
i.e. $G_{{{\m'}}\n'}$ is the standard $D=10$ target space `string-frame' 
 metric. 
If one  formally sets $\t^\a =0$,  only the 
 usual  \sm  type  couplings to the 
  NS-NS fields $G_{{{\m'}}\n'}$ and  $B_{{{\m'}}\n'}$  survive in the action 
(one should   also  add the  standard dilaton 
coupling term  $\sqrt {g^{(2)}} R^{(2)}\p(x)$).
The RR fields  $A_{{\m'}}$ and  $A_{{{\m'}}\n'\l'}$  have fermionic
 couplings which we would like to determine.

The relation  between the $D=11$ and $D=10$ supervielbeins that generalises
the vielbein  relation  in 
\comp\ is   \refs{\duf,\achu}
\eqn\gen{  \h E^A_M = e^{- {1\ov 3}\P} 
\pmatrix{E^{m'}_{M'}  & e^\P A_{M' }  &  E^a_{M'} + A_{M'} \chi^a \cr
0 & e^\P  &   \chi^a \cr}  , }
where $\Phi=\p + O(\theta)$, $\chi^a$ and $A_{M'}$ are  the 
corresponding $D=10$ superfields ($M'=(\m',\a),$ $ A=(m',11,a)$).
Again, the  rescaling by  $e^{-{1\ov 3} \Phi}$ is  needed  to get the 
NS-NS terms in the action   without 
 dilaton factors \achu.\foot{The importance of the rescaling of all of the components of the supervielbein by $
e^{-{1\ov 3} \p}$ is evident, e.g.,  from considering
$\h E^m_i = \del_i x^\m  \h E^m_\m  +  \del_i \t^\a  \h E^m_\a  \to 
e^{-{1\ov 3} \p} ( e^{m'}_{{\m'}} \del_{i'} x^{{\m'}} - \ha i \bar \t \g^{m'} \del_{i'}  \t + ...). $
If $ \hat E^m_\a$ was not rescaled,   $\del x$ and $\bar \t  \del \t$ terms
would have  the relative factor of $e^{-{1\ov 3} \p}$.}
In what follows we shall ignore the dependence  of the action  on the 
 gravitino  and dilatino fields. 
Setting $\psi_\m=0$   and 
adding  the 
extra factors of $e^{-{1\ov 3} \p}$ in front of the expansions in \sup \ 
 we find 
the following expressions for  the relevant components of the  $D=11$ supervielbein
in terms of the  bosonic  fields of the $D=10$ supergravity (cf. \comp) 
\eqn\supe{\h E^{m'}_{{\m'}} =  e^{-{1\ov 3} \p}  {e}^{m'}_{{\m'}} 
 + O(\t^2)\equiv  \h e^{m'}_{{\m'}}  + O(\t^2)  ,
\ \ \  \h E^{11}_{{y}}= e^{{2\ov 3} \p} + O(\t^2) ,  }
$$\h E^{m'}_\a = \ha i e^{-{1\ov 3} \p}(\bar \t \g^{m'})_\a +  O(\t^3), \ \ \ 
\ \h E^a_\a =e^{-{1\ov 3} \p}  \delta^a_\a +  O(\t^2) ,  $$
$$
\h E^a_{{\m'}} = e^{-{1\ov 3} \p}[ \four  
\h \omega_{\m' mn} (\h {e}) \g^{mn}\t + 
 \h F_{\k\n\l\r }  
 \h e^{\k}_{{m'}} \h e^{\n}_{n'} \h e^{\l}_{p'} 
 \h e^{\r}_{q'} \h e^{s'}_{\m'} T_{s'}^{\ m'n'p'q'} \t]^a  + ...
$$
\eqn\fff{ = e^{-{1\ov 3} \p} V^a_{\m'} + ...,  \ \ \  \     
 V^a_{\m'} \equiv  e^\p (\four F_{\m'\n'} \g^{\n'} \g^{11}\t
+    F_{ \l'\r' \k' \s'}  T_{\m'}^{\ \l'\r' \k' \s'} \t)^a  \ .           
}
$F_{\m'\n'} = 2\del_{[\m'} A_{\n']}$ originated from
$\h \omega_{\m' m'11} $ and 
$ F_{ \l'\r' \k' \s'} =  4\del_{[\l'} A_{\r' \k' \s']}  
- 12 A_{[\l'} \del_{\r'}B_{\k' \s']}$
(note that $\hat e^y_{m'}$ $  =-  e^{{1\ov 3} \p}  A_{m'}$).
The  indices in $V^a_{\m'}$ are now  contracted using $e^{m'}_{\m'}$.

For simplicity we  shall 
consider only  part of    RR coupling 
terms which  come  from the  WZ-type  term 
  $\third \ep^{ijk} \h E^A_i\h E^B_j \h E^C_k A_{CBA}$ 
 in \mem. 
Using \aaa\ one  has 
\eqn\wwz{-\third \ep^{ijk} \h E^A_i\h E^B_j \h E^C_k A_{CBA}
= \four \ep^{ijk} \del_i Z^M \h E^m_M \del_j Z^N  \h E^n_N 
 \del_k Z^K \h E^a_K  (\bar \t \g_{mn})_a
+ ... \ . }
Under the double dimensional reduction 
a non-vanishing contribution comes from
the term with $j$ (or $i$) equal to 3 and $N$ (or $M$) equal to 11, 
\eqn\frf{ -\third \ep^{ijk} \h E^A_i\h E^B_j \h E^C_k A_{CBA}
= - \ha \ep^{i'k'} \h E^{11}_y \del_{i'} Z^{M'} \h E^{m'}_{M'}   
 \del_{k'} Z^{K'} \h E^a_{K'}  (\bar \t \g_{{m'}} \g_{11})_a
+ ... \ }
$$ =  \ha \ep^{i'k'} 
(e^{m'}_{{\m'}} \del_{i'} x^{{\m'}} - \ha i \bar \t \g^{m'}
 \del_{i'} \t ) (\bar \t \g_{m'} \g_{11} \del_{k'} \t
+  \bar \t \g_{m'} \g_{11} V_{\n'} \del_{k'} x^{\n'}) + ... \  . 
$$
Notice that the overall powers   of $e^\p$ have cancelled out
except for the one in the  RR  `vertex' $V_{\n'}$. 
The   $O(\t^2)$ part of the  RR coupling term is thus (see \deff,\fff)
\eqn\rra{  e^\p \ep^{i'k'} \del_{i'} x^{{\m'}}  \del_{k'} x^{\n'}
 \ \bar \t  \big[  {{\textstyle{1\over 8}}}  F_{\m'\l'}  \g_{\n'} \g^{\l'}
+     
{{\textstyle{1\over 288}}}F_{ \l'\r' \k' \s'}  \g_{11}  \g_{\n'} (\g^{\l'\r' \k' \s'}_{\ \ \ \ \ \ \ \ \m'} - 8 \g^{ \l'\r'\k' }\delta^{\s'}_{\m'}) \big]\t     . }
In addition,  there is a similar parity-even  RR coupling  contained
in the first $\h E^2$ term in \mem\ and 
coming 
 from higher-order terms in the expansion of the superfield 
 $\h E^{m}_{M}$.

3.  The presence of the $e^\p$ factor in \rra\ 
implies  that the leading-order 
conformal invariance (or $\k$-invariance) equation will  
have the expected form $R +  D^2 \p + (dB)^2 + e^{2\p} (F_2^2 + F^2_4)+ ...  =0$, 
   equivalent  to the 
$D=10$ supergravity equation expressed in terms of the string-frame metric.
One also concludes that  
   higher  powers  $F^k$ of  RR fields will  appear in the $D=10$ type II 
superstring effective action  being multiplied by 
$e^{(k+n -2)\p}$, where 
$n$ is a number of  string loops (genus of the  world sheet).

\bigskip

The author  acknowledges  the support of PPARC and 
ECC grant SC1$^*$-CT92-0789.


\vfill\eject
  \listrefs
\vfill\eject
\end